\documentclass[12pt,thmsa]{article}
\usepackage{sw20lart}

\input{tcilatex}
\begin{document}

\title{On a new Bell test with photons seemingly violating quantum predictions}
\author{Emilio Santos \\
Departmento de F\'{i}sica, Universidad de Cantabria. Santander. Spain}
\maketitle

\begin{abstract}
A disagreement of the empirical results with quantum mechanical predictions
is pointed out in the experiment by B. G. Christensen et al. The anomaly is
quite similar to one present in the analogous experiment by M. Giustina et
al.
\end{abstract}

A new experiment involving photon pairs has been performed violating a Bell
inequality free of the ``fair sampling'' assumption\cite{1}. The purpose of
this note is to point out that one of the correlations measured in the
experiment does not agree with the quantum mechanical prediction. The
anomaly is quite similar to the one present in the experiment of Giustina et
al.\cite{2} (see comment on it\cite{3})

In the experiment\cite{1} entangled (not maximally) photon pairs are
produced by parametric down conversion in such a way that the quantum state
of the pair may be represented by 
\begin{equation}
\mid \Psi \rangle =\frac{1}{\sqrt{1+r^{2}}}\left( \mid HH\rangle +r\mid
VV\rangle \right) ,  \label{1}
\end{equation}
with $r=0.26$ and $H(V)$ denotes horizontal (vertical) polarization of
Alice's and Bob's photons. The quantum prediction for the probability of a
coincidence count with the measuring devices set at angles $\alpha $ and $%
\beta $ is 
\begin{equation}
p_{AB}\left( \alpha ,\beta \right) =\frac{1}{1+r^{2}}\left( \sin \alpha \sin
\beta +r\cos \alpha \cos \beta \right) ^{2}.  \label{2}
\end{equation}

In the experiment four correlations, $C\left( \alpha _{i},\beta _{j}\right)
, $ $i,j=1,2$, were measured at each of the four settings described by the
angles $a=3.8{{}^{o}},$ $a^{\prime }=-25.2{{}^{o}},$ $b=-3.8{{}^{o}},$ $%
b^{\prime }=25.2{{}^{o}.}$ Also two single counts, $S\left( \alpha
_{1}\right) ,S\left( \beta _{2}\right) $ were measured. As a result of the
data the authors report a violation of the measured (Clauser-Horne) Bell
inequality by $7-\sigma $.

In order to make a comparison between the results of the experiment and the
predictions of quantum mechanics it is necessary to estimate the number, $N$%
, of entangled photon pairs produced for every setting of the polarization
analyzers. But before doing that we must normalize to the same number of
trials per setting (see Table 1 of Ref.\cite{1}). Thus we divide the
measured number of counts per setting by the number of trials for that
setting and multiply times 28,000,000. This gives the numbers shown in the
second row of Table 1. After that we estimate the number $N$ of photon pairs
produced in 28 million trials. Actually we need only the product of $N$
times the product of the detection efficiencies, $\eta _{1}\eta _{2},$ which
is obtained from the condition
\[
\sum_{j=1}^{4}\left( E_{j}-N\eta _{1}\eta _{2}Q_{j}\right) ^{2}=\min
.\Rightarrow N\eta _{1}\eta _{2}=\frac{\sum_{j=1}^{4}Q_{j}E_{j}}{%
\sum_{j=1}^{4}Q_{j}^{2}},
\]
where $E_{j}$ are the numbers in the second row of Table 1 and $Q_{j}$ are
the quantum predictions obtained from eq.$\left( \ref{2}\right) $ for the
four combinations of $a,a^{\prime }$ with $b,b^{\prime }$. I get $N\eta
_{1}\eta _{2}=518,037.$ Mutiplication of this number times the quantum
predictions for the four settings gives the last row of Table 1.

\textbf{Table 1}. Comparison between the results of the experiment and the
quantum prediction for the coincidence counts. Numbers in the first (second,
third) row correspond to the raw data of the experiment (corrected data,
quantum prediction) .

$
\begin{array}{lllll}
Settings & C(a,b) & C(a,b^{\prime }) & C(a^{\prime }b) & C(a^{\prime
},b^{\prime }) \\ 
Experiment & 29,173 & 34,145 & 34,473 & 1,862 \\ 
Exper.corrected & 30,008 & 33,721 & 34,687 & 1,867 \\ 
Quantum & 31,419 & 33,553 & 33,553 & 484
\end{array}
$

The disagreement between the empirical data and the quantum predictions in
the former three columns might be explained by experimental errors. Indeed
they are a few times larger than the expected statistical uncertainties. In
contrast there is a dramatic difference in the latter correlation $C\left(
a^{\prime },b^{\prime }\right) ,$ where the empirical result is almost four
times the quantum prediction.

The commented experiment is not the first one exhibiting the anomaly\cite{3}%
. For these reasons a careful investigation of the matter is worth while.


\begin{thebibliography}{9}
\bibitem{1}  B. G. Christensen et al., Arxiv/1306.5772.

\bibitem{2}  M. Giustina et al., \textit{Nature} \textbf{497}, 227-230
(2013). Arxiv/1212.0533.

\bibitem{3}  E. Santos, Arxiv/1305.6876
\end{thebibliography}
\end{document}